\documentstyle[astrobib,psfig]{mnv2}
\def\gs{\mathrel{\raise0.35ex\hbox{$\scriptstyle >$}\kern-0.6em 
les
\lower0.40ex\hbox{{$\scriptstyle \sim$}}}}
\def\ls{\mathrel{\raise0.35ex\hbox{$\scriptstyle <$}\kern-0.6em 
ggles
\lower0.40ex\hbox{{$\scriptstyle \sim$}}}}
\def\ltorder{
\mathrel{\raise.3ex\hbox{$<$}\mkern-14mu\lower0.6ex\hbox{$\sim$}}
}
\def\gtorder{
\mathrel{\raise.3ex\hbox{$>$}\mkern-14mu\lower0.6ex\hbox{$\sim$}}
}
\def\msun{\>{\rm M_{\odot}}}

\input psfig

\title[Quasar outflows and the formation of dwarf galaxies]
{Quasar outflows and the formation of dwarf galaxies}

\author[Priyamvada Natarajan, Steinn Sigurdsson \& Joseph Silk]
{Priyamvada Natarajan,$^1$, Steinn Sigurdsson$^1$ and Joseph
Silk$^{1,2,3}$\\
$^1$Institute of Astronomy, Madingley Road, Cambridge CB3 0HA, U. K.\\
$^2$Department of Astronomy and Physics, Univ. of California at 
Berkeley, Berkeley, CA 97420, U. S. A.\\
$^3$Center for Particle Astrophysics, Univ. of California at
Berkeley, Berkeley, CA 97420, U. S. A.}

\begin{document}
\label{firstpage}
\maketitle

\begin{abstract}
In this paper we propose a scenario for the formation of a population
of baryon rich, dark matter deficient dwarf galaxies at high redshift
from the mass swept out in the Inter-Galactic Medium (IGM) by
energetic outflows from luminous quasars. We predict the intrinsic
properties of these galaxies, and examine the prospects for their
observational detection in the optical, X-ray and radio
wavebands. Detectable thermal Sunyaev-Zeldovich decrements on
arc-minute scales in the cosmic microwave background radiation maps
are expected during the shock-heated expanding phase from these hot
bubbles. We conclude that the optimal detection strategy for these
dwarfs is via narrow-band Lyman-$\alpha$ imaging of regions around
high redshift quasars. A scaled down (in the energetics) version of
the same model is speculated upon as a possible mechanism for forming
pre-galactic globular clusters.
\end{abstract}

\begin{keywords}
\end{keywords} 

\section{Introduction}

Galaxy formation is a complex physical process involving several
length, mass and time scales, therefore it is expected to proceed in a
fashion that is modulated by the properties of the local environment
in which it occurs. In the context of a hierarchical picture for
structure formation in a cold dark-matter dominated Universe, most
massive galaxies are thought to assemble fairly recently (at z $\sim$
0.5 - 1) from the agglomeration of sub-clumps that have collapsed at
higher redshifts (\citeNP{sdmw91}; \citeNP{kauffmann93};
\citeNP{baugh97}).  Recent observations from the Hubble Space
Telescope ({\it HST}) and ground-based telescopes find that a
significant fraction of the stellar mass as traced by blue light, that
constitute galaxies is in place by redshift $\sim$ 1.5 - 2.0
(\citeNP{madau96}; \citeNP{lilly96}). The other cosmological objects
in place by these redshifts are quasars; and given the high bolometric
luminosities of high redshift quasars it is natural to expect that
they probably play an important role in locally modulating galaxy
formation (\citeNP{efsta88}; \citeNP{babul91}; \citeNP{babul92}; 
\citeNP{joemjr98}).

We propose that quasars effect galaxy formation in their vicinity via
energetic outflows. The outflow powered by the mechanical luminosity
of the quasar shocks and sweeps out mass from the IGM into a (thin)
shell that propagates out on scales of the order of a few hundred kpc
or so. The swept up matter expands adiabatically till a gravitational
instability of the shell causes it to fragment, yielding clumps with a
typical mass-scale, $\sim\,{10^9}\,\msun$ corresponding to the mass of
dwarf galaxy size objects, akin to the explosion-driven galaxy
formation scenarios proposed by \citeN{mckee77}, \citeN{ost81} and
\citeN{ost83}. The propagation of these cosmological blast-waves has
been computed by \citeN{ost81}, \citeN{bert85}, \citeN{vish85},
\citeN{carr85}, \citeN{teg93} in slightly different contexts, that of
explosion induced structure formation models or the role of supernovae
winds. In recent papers, \citeN{voit96} has examined the effects of
quasar blown bubbles on the structure of the Inter-Galactic Medium,
here we focus on the fate of the gas in these bubbles. We consider the
basic scaling relations for our proposed model in the second section
of this paper. In Section 3, we outline the dynamics of quasar
outflows and the fate of the cold dense gas deposited in shells around
the quasar, when gravitational instability causes mass clumps to form
at the scale of dwarf galaxies. In Section 4, the predicted properties
and prospects for detection of this population of dwarfs and the
outflow in the optical and X-ray wave-bands are explored. We then note
that the high-redshift blue clumps detected by {\it HST} reported by
\citeN{pascarelle96} and the galaxies detected in the Lyman-$\alpha$
by \citeN{hu96} are potential candidates (Section 5). Before
concluding, we speculate on the consequences of a scaled-down version
of the same scenario for the formation of the pre-galactic globular
cluster population.

\section{Quasar outflows}

The total energy budget available from a quasar to power the
mechanical luminosity of the outflow is determined by the mass accreted by a
black hole of mass $M_{\rm bh}$ modulo two efficiency factors: one
that determines the conversion of rest mass into the bolometric
luminosity $\epsilon\,(\sim 0.1)$, and the efficiency of conversion of
the bolometric luminosity into mechanical luminosity parameterized as
$\epsilon_{\rm L}\,(\sim 0.5)$.  Bolometric luminosities of bright
quasars at redshifts $z \gtorder 2$ are as high as $\sim 10^{47}\,
{\rm erg\, s^{-1}}$ or larger.  A typical outflow can therefore sweep
up mass $M_{\rm IGM}$ from the IGM:
\begin{eqnarray}
{M_{\rm IGM}}\,&\sim&\,\nonumber{2\,\times\,{10^{12}}}\,({\frac
{L_{\rm QSO}}{10^{47}\,{\rm erg\,s^{-1}}}})\,({\frac
{\delta t}{{5\,\times\,10^7}\,{\rm yr}}})\,\\&\times&({\frac
{v_{\rm flow}}{3000\,{\rm km\,s^{-1}}}})^{-2}\,\msun,
\end{eqnarray}
where $\delta t$ is the duration of the wind phase/the lifetime of the
quasar, $v_{\rm flow}$ the outflow velocity and $L_{\rm QSO}$ the
mechanical luminosity of the quasar. The outflow extends to:
\begin{eqnarray}
R\,&\sim&\,\nonumber {500}\,({\frac {{\epsilon}\,\epsilon_L}{0.05}})^{1/3}
\,({\frac {M_{\rm bh}}{10^{9}\,\msun}})^{1/3}\,({\frac
{f_Q}{0.01}})^{-1/3}\\ &\times& ({\frac {v_{\rm flow}/c}{0.01}})^{-2/3}\,{\rm kpc}.
\end{eqnarray}
where $f_Q$ is the local baryon fraction in units of 0.01. Note that
if the density of the ambient medium is lower, then the outflow
expands further out on to larger scales. All the numbers quoted in
this work have been computed for ${H_0}\,=\,50\,{\rm
km\,s^{-1}\,Mpc}$; ${\Omega_0}\,=\,1.0$ and $\Lambda\,=\,0$.

\section{Outflow dynamics}

Below, we summarize the dynamical sequence before estimating the
relevant numbers. In this treatment, we ignore the detailed
microphysics of the outflow from the BH, but assume that the flow is a
mildly relativistic and directed one involving only a small mass at
the base of the outflow. We also ignore the time evolution of the
initial stages of the outflow and consider the dynamics of the outflow
only at the epoch when the swept up mass is comparable to or greater
than the mass of the BH $M_{\rm bh}$.

During the lifetime of the quasar ($t_{\rm Q}\,\sim\,{5 \times 10^7}$
yr), the outflow expands freely into the IGM, shock heating material
locally to high-temperatures $\sim\,{10^8}$ K. During the hot phase,
we expect X-ray emission from the shocked material. For quasars that
switch on at high redshifts ($z\,>\,7$) inverse Compton cooling will
dominate but at low redshifts, the shocks expand adiabatically and
cool radiatively. As the mass accumulated around the shock becomes
dynamically significant a reverse shock propagates thermalizing the
gas that is concentrated along a thin shell. This shell subsequently
evolves adiabatically, akin to supernova remnants until the age of the
bubble becomes comparable to the total elapsed Hubble time at that
redshift, at which juncture cosmological effects alter the structure
of the outflow and its growth.

Outflows originating after z $\sim$ 7, with energies greater than
10$^{57}$ erg remain adiabatic to low redshifts (\citeNP{bert85}).
The assumption of adiabatic expansion remains valid (the post-shock
density being roughly four times the density of the ambient IGM; and
the radius R(t) of the bubble scaling as t$^{2/5}$ once the QSO shuts
off) until the shell of mass swept up at the shock front becomes
self-gravitating.  In this treatment we focus on the dynamics of the
outflows during this radiative cooling dominated stage. They are
well-approximated by the family of self-similar Sedov-Taylor blast-wave
solutions (\citeNP{mckee77}; \citeNP{bert85}; \citeNP{voit94};
\citeNP{voit96}) at this epoch. The swept up mass becomes dynamically
important at a time $t$ after the commencement of the outflow:
\begin{eqnarray}
t\,&\sim&\,5\,\times\,10^7\,(1+z)^{-3/2}\,{\Omega_{\rm
IGM}^{-1/2}}\,(\frac {\dot M}{4 \times 10^4 \,M_{\odot}\,{\rm
yr^{-1}}})^{1/2}\\ \nonumber &\times& (\frac {v_{\rm flow}}{3000\,{\rm km
s^{-1}}})^{-1/2}\,{\rm yr},
\end{eqnarray}
where ${\dot M}$ is the mean outflow rate (i.e. ${M_{\rm
IGM}}/{t_{\rm Q}})$ in units of $4 \times 10^4$ M$_{\odot}\,{\rm
yr^{-1}}$ and ${\Omega_{\rm IGM}}$ is the mass density of the IGM
(where ${\Omega_{\rm IGM}}\,<<\,\Omega_0$), at which time the bubble radius is:
\begin{eqnarray}
R\,&\sim&\,{400}\,(1+z)^{-3/2}\,{\Omega_{\rm
IGM}^{-1/2}}\,(\frac {\dot M}{4 \times 10^4\,M_{\odot}\,{\rm
yr^{-1}}})^{1/2}\\ \nonumber &\times&(\frac {v_{\rm flow}}{3000\,{\rm km
s^{-1}}})^{-1/2}\,{\rm kpc}.
\end{eqnarray} 
Eventually, as the age of the bubble approaches the local Hubble time,
the cold shell of accumulated material fragments due to gravitational
instability (although in principle, while Rayleigh-Taylor
instabilities could set in, they do not result in the formation of
gravitationally bound fragments). The boundary conditions immediately
prior to fragmentation can be estimated following the treatment of
\citeN{ost81} and \citeN{voit94}. The radiative cooling time
$\tau_{\rm cool}$ is:
\begin{eqnarray}
\tau_{\rm cool}\,=\,{1.2\,\times\,{10^8}}\,{(\frac T
{10^5\,K})^{11/5}}\,{(1+z)^{-3}}\,\,{\rm yr}. 
\end{eqnarray}
Equating the radiative cooling time to the age of the bubble, an
expression for the post-shock velocity $v_{\rm cool}$ can be obtained
(\citeNP{ost81}) in terms of the total energy injected into the
outflow,
\begin{eqnarray}
v_{\rm cool}\,\sim\,{150}\,({\frac {L_{\rm QSO}}{10^{47}\,{\rm erg
s^{-1}}}})^{1/20}\,
({\frac {\delta t}{5\,\times\,10^{7}\,{\rm yr}}})^{1/20}\,(1 +
z)^{1/3}\,\,{\rm km\,s^{-1}}.
\end{eqnarray}
The shell temperature can be related to the total swept up mass as follows,
\begin{eqnarray}
T_{\rm shell}\,=\,1.4\,\times\,10^5\,{(\frac {M_{\rm sweep}}
{10^{12}\,M_{\odot}})^{2/5}}\,{(1 + z)}^{3/5}\,\,K
\end{eqnarray}
The maximum mass that can cool during the radiative cooling phase is
$10^{12}\,M_{\odot}$ and fragments less than $10^{9}\,M_{\odot}$ are
not Jeans unstable at this epoch, thereby naturally providing typical
mass-scales. Fragmentation on a scale $\xi$ can therefore occur if a
small disk of radius $\xi$ and mass $\sim\,{{M_{\rm
sweep}}\,{\xi^2/R^2}}$ extracted from the shell has a gravitational
collapse time that is less than the sound-crossing time along $2\xi$.
The instability therefore imprints a characteristic mass scale,
\begin{eqnarray}
M_{\rm frag}\,\sim\,{5\,\times\,10^9}\,({\frac {E}{10^{63}\,{\rm erg}}})^{-3/10}\,(1 +
z)^{1/5}\,\,{\rm M_{\odot}},
\end{eqnarray}
which is roughly $10^{9 - 10}\,\msun$ (\citeNP{mckee77},
\citeNP{ost81}) for the numbers in equations (1) and (2); leading to
the formation of a group of dwarf galaxies around the quasar from the
total swept up mass. While these clumps are not generally bound to
each other, they are highly clustered both spatially and in velocity
space. Note here that lower energy outflows imply a high fragment mass
although a smaller value of the gas mass is swept up initially,
therefore weaker outflows from lower luminosity QSOs or super-winds
(\citeN{BM88}; \citeN{heck87}) could be important for the formation
of dark matter poor galaxies at lower redshifts.

The outflow material can suffer one of three fates: it may stall and
fall back, if the mass swept up is large and the central potential
well is deep (eg. if the quasar is at the centre of a proto--cluster);
the outflow may coast to infinity (in practice till it matches on to
the Hubble flow); or, it may escape the central object hosting the
quasar but be bound to the proto--cluster, yet not fall back to the
centre on radial orbits if the proto--cluster potential is strongly
non--spherical. The condition for fall back can be estimated using a
density profile $\rho(r)_{\rm DM}$ to describe the dark matter profile
around the quasar. The fragments coast on only if $v_{\rm flow}$
exceeds the local escape velocity $v_{\rm esc}$ of the total potential
at a given distance $r$ from the nucleus:
\begin{eqnarray}
v_{\rm esc}\,\sim\,3000\,({\frac {M(r)} {10^{15} M_{\odot}}})^{1/2}\,
({\frac {r} {1\,{\rm Mpc}}})^{-1/2}\,\,{\rm km\,s^{-1}}
\end{eqnarray}

\subsection{Traditional scenarios for the formation of dwarfs}

Dwarf galaxies can form directly from the collapse of cosmological
density fluctuations at high redshift (\citeNP{ikeuchi87}), in which
case we expect them to be dark matter dominated, specially since the
first generation of star formation may unbind most of the gas in the
disk (\citeNP{dekel86}), unless the disk formed from a high angular
momentum perturbation, has low surface density with sparse star
formation, in which case we would see the high z counter-part of a gas
rich object with a dark matter dominated rotation curve at low z
(\citeNP{deblok96}).

\citeN{babul92} however, have argued that the cosmological collapse of
dwarf galaxy scale masses is inhibited until $z\,\leq\,1$ due to the
photo-ionization of the IGM by the meta-galactic UV
radiation. Therefore, their model and subsequent work by
\citeN{babul96} predict a formation scenario for low-mass galaxies at
recent epochs that fade after a burst of star-formation at
$0.5\,\leq\,z\,\leq\,1$. These resultant low luminosity, low surface
brightness objects are claimed to dominate the blue number counts at
faint magnitudes in redshift surveys (\citeNP{ellis97}) but cannot however
 be reconciled with the red counts (cf. \citeNP{bouwens}).

Interactions of large, gas rich spirals may also lead to the formation
of gas rich, dark matter poor dwarf galaxies by the fragmentation of
extended or unbound tidal tails (\citeNP{dubin96}).  Such dwarfs would
form in small numbers near major mergers and are likely to fade
rapidly after the initial burst of star formation, and would only be
seen in the close vicinity of recently merged spirals.

The population of dwarf galaxies postulated here is distinguished by
forming predominantly at high redshift, in large groups spread in an
approximate sheet geometry over $\sim$ Mpc scales. These dark matter
deficient objects are characterized by low amplitude rotation curves
and low central velocity dispersions. Since they form from the
fragmentation of a thin shell, the clumps have low specific angular
momentum and are hence likely to collapse into relatively compact
structures, with a brief star forming phase until supernovae remove or
disperse the bulk of the remaining gas (\citeNP{dekel86};
\citeNP{cmjr86}). These dwarfs would form in small numbers ($\sim 1 $
per $L_*$ galaxy, at high $z$ only, but may be observationally of
interest due to their strong clustering and association with nearby
bright quasars.

\section{Predicted properties of the population}

The dwarf galaxies that form on the cooling of these fragments are
expected to have the following properties:
\begin{itemize}

\item{are baryon-rich and dark matter deficient, so the typical
rotation curve will have a much lower amplitude than say for a local
dwarf with a comparable luminous mass; hence these dwarf galaxies are
expected to have disk-dominated rotation curves.}

\item{Since the swept up material is principally from the IGM at high
redshift, that is not very enriched, they are expected to be
metal-poor, with $\langle Z \rangle \, \sim
10^{-2}\,Z_{\odot}$,}

\item{since they do not contain significant amounts of dark-matter, we
expect that the first few supernovae (SN) can entirely disrupt the
remaining gas disk, truncating star formation, and therefore the
surface number density of these galaxies per square degree is likely
to decline rapidly at lower redshifts, as the galaxies fade after the
initial burst of star formation.}

\item{These dwarfs are expected to be highly clustered as they form in
a sheet around the quasar. Therefore, any line of sight to a background
quasar through a group of nascent dwarfs, may intersect 
many individual members with narrow velocity spacing, producing a
characteristic signature in the absorption line profiles. 
Each absorption line would also be relatively narrow compared
to lines of the same column density at low redshifts, as the internal
velocity dispersion of these dark matter deficient galaxies is lower,
and as the associated gravitational potentials are fairly shallow, while 
their baryonic fraction is extremely high.}

\item{the gas that settles down into the disk in these systems is
likely to form compact, sub-critical disks (sub-$L_*$) and hence
star-formation is likely to ensue in intense bursts, in
isolated knots,}

\item{since they form from gas clumps with low specific angular
momentum, the gas can be funnelled in to the center very efficiently
to form a compact central object. Therefore, these dwarfs are likely
to harbor weak AGN, with narrow-emission lines - $M_{BH} \ltorder 10^6 \msun$,
 with luminosities less than $10^{44}$ erg/s.}

\item{Given the typical gas masses of these dwarfs, using a standard
Schmidt law for the conversion of gas into stars we estimate star
formation rates of the order of 30 - 60 M$_{\odot}\,{\rm yr^{-1}}$.}

\item{If these dwarfs/pre-fragmentation gas clumps remain weakly bound
to the proto-cluster potential, they may form stars and avoid falling
back to the central galaxy if strong inhomogeneities in the potential
scatter them onto tangential orbits in the proto--cluster. In such
cases, it is conceivable that a trace of this population might be
observed even in the local neighbourhood, despite the fact that the tidal
forces in the cluster would likely have disrupted the dwarf galaxies
themselves. With $\sim 10^{10}$ stars per dwarf, we might expect a few
bright planetary nebulae (PN) per (disrupted) dwarf galaxy. These PN
would contribute to the apparent extra-galactic background of PN
detected in local clusters (\citeNP{ciar97}, \citeNP{ferg97},
\citeNP{ciar89}). These PN would be distinguishable via their clumpy,
locally flat density distribution, with small velocity separation;
since we expect the internal dispersion of the dwarfs to be
small. Even after disruption the stars and hence PN tracers of the
stellar population would stay as moving groups, shearing out only over
several cluster dynamical times. It is possible that deep imaging
would reveal an excess of fainter PN near individual bright PN,
thereby, tracing the stellar population of the dwarf clump of stars.}

\item{these dwarfs are expected to be detected in the radio - via their
low-level emission from the weak central AGN; in the optical - as 
a consequence of a burst of star-formation; on being significantly 
magnified as a result of gravitational lensing; or via narrow-band 
Lyman-$\alpha$ imaging by detecting either the presence of the warm, 
diffuse gas or emission from the dwarfs undergoing their first episode
of star-formation.}

\end{itemize}

Despite the fact that the covering fraction along the line of sight to
a background QSO of these systems is small, being gas rich, they might
well be detectable in absorption, perhaps explaining the curious fact
that the Lyman-limit systems detected by \citeNP{steidel96};
\citeNP{steidel97} using the color selection criteria at high redshift
($z\,\geq\,3$) are {\bf not} the optical counter-parts of the
absorbers seen in quasar spectra in the same redshift range (unlike
the case at low redshifts where such a correspondence has been
well-established) - pointing to the fact that the absorption
at high-redshifts is probably being produced by gas-rich galaxies like
these dwarfs that are either yet to commence their first episode of
star formation or are intrinsically very faint.

\subsection{X-ray emission predictions}

In the following calculation of the X-ray emission, we assume that a
virialized cluster has {\bf not} assembled around the quasar (although
the quasar might seed an over-dense region), therefore the X-ray
emission arises primarily due to shock-heating of the outflow. During
the early adiabatic expansion phase of the hot, shocked layer of the
bubble, high temperatures are expected $\sim\,10^8$ K. Therefore,
prior to cooling, we expect to see X-ray emission from the hot
bubble. We model the density profile of the hot bubble with a constant
density out to radius $(R - w)$, where $R \sim 1$ Mpc and $w$ the
width of the shell. Using the standard Hugonoit jump conditions across
a strong shock, the shell density at the outer edge is four times the
central value;
\begin{eqnarray}
n_e(r)\,=\,n_0\,\,\,\,\,\,\,\,0 \leq r \leq (R-w)\,;\\
n_e(r)\,=\,4 n_0\,\,\,\,\,\,\,(R-w) \leq r \leq R. 
\end{eqnarray}
Outside $R$, the density is assumed to be that of the ambient
IGM. This profile shape is motivated by the density profile found by
\citeN{bert85} for the adiabatic cosmological detonation wave
propagation solution. The thickness $w$ is estimated to be,
$w\,\sim\,{t_{\rm cool}}\,{v_{\rm shock}}\,\sim\,100$ kpc. The central
density $n_0$ is computed by requiring that the enclosed mass within
radius $R$ is equal to the total mass ($M_{\rm
flow}\,\sim\,10^{12}\,M_{\odot}$) swept up by the outflow, which gives
$n_0\,\sim\,{10^{-2}}$ cm$^{-3}$. 
The total emissivity is given by:
\begin{eqnarray}
\epsilon(r)\,=\,{1.4 \times 10^{-27}}\,{T_e^{\frac 1
2}}\,{n_e^2(r)}\,\,\,\,{\rm erg\,\,s^{-1} cm^{-3}};
\end{eqnarray}
and the integrated bolometric X-ray luminosity,
\begin{eqnarray}
L_{\rm X}\,=\,\int_0^{\rm r_{max}}{\epsilon(r)}\,{4 \pi
r^2}\,{g(\theta_{\rm open})}\,\,dr\,\,\,\,\,\,{\rm erg\,\,s^{-1}}
\end{eqnarray}
where ${g(\theta_{\rm open})}$ takes into account the opening angle of
the outflow. The total X-ray flux from a bubble at redshift $z$ can
then be easily computed to examine the feasibility of detection. In
Fig. 1, we plot the allowed region for given physical properties
(central gas density $n_0$ and temperature for a significant detection
(5$\sigma$) by ROSAT, ASCA and AXAF at their flux detection limits,
respectively, 10$^{-14}$ erg cm$^{-2}$ s$^{-1}$, 
10$^{-15.5}$ erg cm$^{-2}$ s$^{-1}$ and 10$^{-15}$ erg 
cm$^{-2}$ s$^{-1}$; during a 10$^3$ ks integration.
\begin{figure}
\centerline{
\psfig{figure=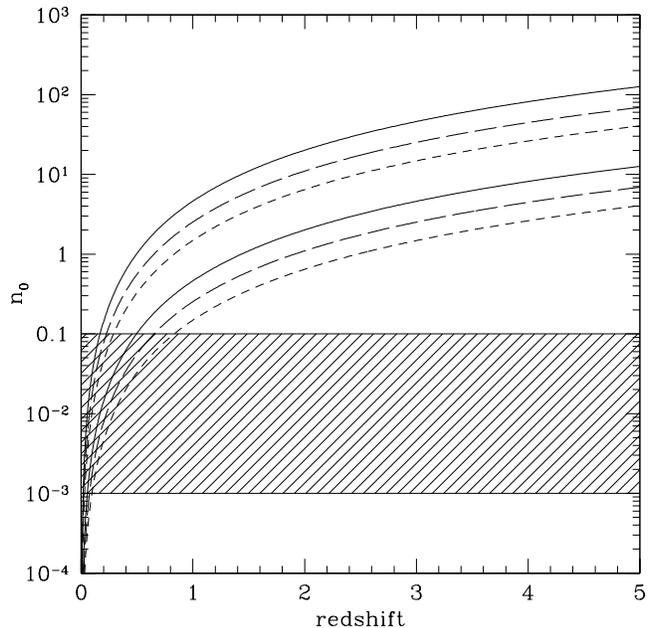,width=0.5\textwidth}}
\caption{The predicted physical properties of the hot gas at the
base of the outflow for a 5$\sigma$ detection at the flux limit of
ROSAT (solid curves); ASCA (long-dashed curves) and AXAF (short-dashed
curves) following a 10$^3$ ks integration. The shaded
area marks the range in the estimated physical properties of the
outflow gas for 2 temperatures $10^8$ K (upper curves) and $10^6$ K
(lower curves).}
\end{figure}
The X-ray luminosity computed here is the integrated bolometric
value, the contribution to either the soft-band or the hard-band
individually are likely to be significantly lower.  Therefore, we see
clearly from Fig. 1, that beyond z = 1, we are unlikely to detect any
direct X-ray emission from these hot bubbles. But, for lower redshift
we see that these sources could have been detected in the current
deep X-ray surveys. Therefore, a systematic search for detected
extended sources that do not correspond to optically detected clusters
in the vicinity of quasars could reveal the presence of these
outflows at $z\,\sim\,1$.

Note that above we have assumed that the density contrast between the
interior of the bubble and the ambient ISM $\Delta {n_e}\,\sim$ an
order of magnitude. If indeed the QSO is seeding a proto-cluster then
this contrast can be much larger, implying a much lower density for
the ambient medium which in turn (as can be easily seen from eqn (2)) leads to
the bubble expanding further out to much larger scales, still getting
shock-heated to temperatures of a few times $10^8$ K. In that case,
the subsequent fragmentation and cooling phases would be delayed.  In
both scenarios, the optical depth $\tau$ has comparable values,
$\tau\,\sim\,{n_e\,R}\,\sim\,{10^{-3}}$, therefore, despite their low
X-ray luminosity, these hot bubbles are expected to produce thermal
Sunyaev-Zeldovich (S-Z) decrements (due to the scattering of the cool
microwave background photons off the hot electrons in the bubble) that
may be detectable,
\begin{eqnarray}
{\frac {\Delta T}{T}}\,=\,({\frac {2\,k\,T_e} {m_e\,c^2}}\,\tau)\,
=\,{3\,\times\,{10^{-5}}}\,({\frac {T_e}{10^8\, K}}).
\end{eqnarray}
The X-ray emission may only be observable over a small range in
redshift, but the temperature decrement may show up at all $z$ as
arc-minute scale patches in the cosmic microwave background.  A
general scenario to explain the detected S-Z decrements
(\citeNP{jones97}) in the absence of assembled, hot clusters at high
redshift from quasar outflows (both a thermal and a kinematic S-Z
component) has been explored by \citeN{pri97b}. 

The energy available for the scale of outflows proposed here is more
than an order of magnitude lower than for the case speculated upon by
\citeN{pri97b}, with both $L_{QSO}$ and $t_Q$ smaller than envisaged
in the earlier paper. They proposed a kinematic S-Z decrement assuming
rapid cooling via turbulent mixing, while the decrement in the case
discussed here is dominated by thermal S-Z effect that persists till
the gas cools to $10^6 \, K$.  A thermal S-Z effect is temporarily
observable while the powering QSO is on, and for a short time
afterwards, despite the fact that the gas is unbound, because
expansion and hence adiabatic cooling is slow in the self-similar
expansion regime. Since it takes the gas, $t\,\sim\, {R/v_s}\,\sim\,3
\times 10^8$ yr (where R is the radius and $v_s$ the sound speed) to
cool, the bubble may stay hot for long enough for a detectable thermal
S-Z decrement.  During the high luminosity phase of the QSO powering
the outflow, the S-Z thermal decrement would likely be unobservable
because of strong radio lobe emission within the outflow. If the radio
lobes fade faster than the gas cools (\citeNP{rees89}), then there may
be a brief period during which this thermal S-Z effect is observable.
In principle, since the frequency dependence of the thermal and
kinematic S-Z are different, their relative contributions can be
observationally tested.

\section{Observational evidence}

Below, we focus on two recent observations that may be consistent 
with our proposed picture for the modulated galaxy formation
around quasars. \citeN{pascarelle96} reported the discovery of a
possible group of sub-galactic clumps at z = 2.39 located within an
arc-minute of a weak radio galaxy 53W002. Lyman-$\alpha$ emission is
detected with equivalent widths of approximately 40 - 50 angstroms,
typical of ionization arising purely from young stars, indicative
of star-formation activity in these sub-clumps with inferred stellar
masses $\leq\,{10^9}\,\msun$ and star-formation rates
$\sim\,{50}\,\msun\,yr^{-1}$. The magnitudes of these clumps range
from $M_V$ of -23 to -18, and if the stellar population is assumed to
be young, removing the point source contribution, their inferred luminosities
range from (0.1 - 1) $L^*$. Five of these candidates have confirmed
redshifts and three contain weak AGN; the redshift distribution is
very narrow $\Delta z\,\sim\,0.01$, consistent with these clumps being
distributed in a sheet geometry around the radio galaxy. Additionally,
\citeN{hu96} have detected 2 objects during a narrow-band
Lyman-$\alpha$ imaging search in a 3.5' X 3.5' region around a z =
4.55 quasar. Both objects are at the same redshift as the quasar.
They are separated enough spatially ($\sim$ 700 kpc) so that the
quasar is unlikely to be the source exciting the observed
Lyman-$\alpha$ emission, and therefore the emission is most probably
originating from the star-formation activity within these galaxies.

\section{Can we form Globular Clusters?}

Here we speculate on the possibility of the formation of globular
clusters via the same mechanism, with an attempt to explain the excess
number density and possible existence of two distinct populations of
globular cluster systems around bright elliptical galaxies. From the
scaling arguments presented above, we see that weaker outflows (simply
scaling down the parameters in this physical picture) might trigger
fragmentation to lower mass clumps, which would stall before escaping
from the quasar host galaxy. These lower mass clumps should form near 
spherical systems with masses $\sim\,10^6$ $M_{\odot}$, since the
mutual tidal torquing of the clumps cannot impart any significant 
spin to them.

Observationally, an excess in the number of GCs associated with bright
central galaxies is detected (\citeNP{whit96}; \citeNP{fgril97}) which
fits in nicely within this outflow picture. It is possible that some
fraction of the population of globular clusters seen around bright
cluster galaxies formed in weaker outflows propagating through the
IGM, and two-phase warm gas subsequently falling back into the central
potential after fragmentation and cooling. The fragments having
insufficient velocity to escape the central potential, would stall and
collapse in the outskirts of the host galaxy of the quasar, forming a
population of nearly co-eval, metal-poor, globular cluster size
stellar clumps. Forming with low angular momentum relative to the
centre of the potential, the population would assume an $r^{-2}$
density profile with highly radially anisotropic orbits. Any
triaxiality of the central potential would isotropise the spatial
distribution of the globular clusters about the galaxy in a few
orbital periods, leaving the observed shallow surface density profile
of the excess globular population. This scenario, therefore predicts
that the velocity distribution of the excess globulars ought to be
highly radially anisotropic. Recent observational studies of the GC
systems around bright ellipticals like M87 (\citeNP{elson96}), and the
ones in the Fornax cluster (\citeNP{fgril97}) seem to reveal a clear
bimodal colour distribution. In several of these instances, the
metal-rich sub-populations are more centrally concentrated than the
metal-poor ones (NGC 472 - \citeNP{geisler96}; NGC 5846 -
\citeNP{fgril97}; M87 - \citeNP{elson96}; NGC 3115 -
\citeNP{elson97}).  Given these detailed observational studies as well
as the theoretical multi-phase collapse model (consisting of 2
episodes of GC formation - a pre-galactic assembly one and a second
galactic phase) recently proposed by \citeN{fbro97} for the in situ
formation of GC systems around ellipticals; our proposed model is
appealing since it provides a physical mechanism to form GCs in the
pre-galactic phase.

\section{Conclusions}

We have presented a scenario for the formation of baryon dominated,
low-mass dwarf galaxies at high redshift in the vicinity of
quasars. The IGM gas mass swept up by the mechanical luminosity of the
outflow from the quasar forms a thin dense shell that expands
adiabatically till the age of this gas bubble approaches the Hubble
time at the given redshift z, at which juncture the shell fragments
due to a Jeans type gravitational instability producing a
characteristic clump mass of 10$^9$ M$_{\odot}$, leading to the
formation of a group of dwarf galaxies in a sheet-like geometry on
roughly an arc-minute scale around the quasar. We predict the
properties of this high surface-brightness, low-mass, high redshift
dwarf population and demonstrate that the optimal detection strategy
would be narrow-band Lyman-$\alpha$ imaging in the regions around high
redshift quasars. Examining the feasibility of detection in the X-ray
during the early adiabatic expansion phase of the hot bubble, we find
that direct X-ray emission is likely to be detectable only from low
redshifts ($z < 1$), while measurable thermal S-Z decrements on
arc-minute scale patches in the cosmic microwave background are
expected from all redshifts. Finally, we argue that a scaled down
version of the same outflow picture could lead to a fragment mass that
is typical of GCs, therefore providing a possible physical mechanism
for the formation of the older metal-poor, pre-galactic GC population.

\section*{Acknowledgments}

We acknowledge Martin Rees for many useful discussions and comments on
the manuscript. We thank Rebecca Elson for valuable discussions. SS
acknowledges the support of the European Union, through a Marie Curie
Individual Fellowship. The research of JS has been supported in part
by grants from NASA and NSF. JS acknowledges with gratitude the
hospitality of the Institut d'Astrophysique de Paris where he is a
Blaise-Pascal Visiting Professor, and the Institute of Astronomy at
Cambridge, where he is a Sackler Visiting Astronomer.

\bibliography{mnrasmnemonic,refs}
\bibliographystyle{mnrasv2}

\end{document}